\documentclass{elsart}
\usepackage{graphicx} 
\begin{document}
\begin{frontmatter} 

\title{Statistical models of diffusion and aggregation for coke formation in a
catalyst pore}
\author{F. D. A. Aar\~ao Reis}
\ead{reis@if.uff.br}           
\address{
Instituto de F\'\i sica, Universidade Federal Fluminense,\\
Avenida Litor\^anea s/n, 24210-340 Niter\'oi RJ, Brazil}
\date{\today}
\maketitle

\begin{abstract}
We simulated models of diffusion and aggregation in long pores of
small widths in order to represent the basic mechanisms of coke deposition in
catalysts' pores. Coke precursors are represented by particles injected at the
pore entrance. Knudsen diffusion, which is usually expected inside
the pores, is modeled by ballistic motion of those particles.
The regime of molecular diffusion is also analyzed via models of lattice random
walks biased along the pores. The aggregation at the
surface or near previously aggregated particles was modeled by different
probabilistic rules, accounting for the possibilities of more compact or more
ramified deposits. In the model of Knudsen diffusion and in some cases
of molecular diffusion, there is an initial regime of uniform deposition along
the pore, after which the deposits acquire an approximately wedge shape, with
the pore plugging near its entrance. After the regime of uniform deposition
and before that of critical pore plugging, the average aggregation position
slowly decreases with the number $N$ of deposited particles approximately as
$N^{-0.25}$. The apparently universal features of deposits generated by
microscopic models are compared with those currently adopted in continuum
models.
\end{abstract}


\end{frontmatter}

\newpage

\section{Introduction}
\label{intro}

Catalyst deactivation is a phenomenon of great economic impact on many
industrial processes. One of the main causes of deactivation is coke formation,
which consists in the deposition of significant amounts of carbonaceous
residues onto the catalyst surface. These deposits reduce the activity of the
catalyst because they block the active sites and distort the porous structure,
thus reducing the diffusivities of the
reactants~\cite{richardson,buttbook}. The importance of this
phenomenon motivated several theoretical
studies~\cite{froment,mann,bartholomew}, with the deactivation process being
modeled at three different levels~\cite{froment}: the active site, the
catalyst particle (or pellet) and the reactor. The studies at the particle
level are strategic for the design of new catalysts.

Although the morphology of intra-particle coke deposits is essential to
understand catalyst deactivation, it is usually oversimplified
in theoretical models. Symmetrical coke profiles along the pores are frequently
adopted, even when a realistic pore connectivity is considered for representing
the structure of the catalyst. In most works, uniform (flat) coke profiles along
the pore are considered~\cite{beyne,muhammad}.
On the other hand, nearly 20 years ago, Mann and co-workers introduced
models in which the coke deposits inside the pores grew with a wedge shape of
constant slope, with a higher concentration at the pore
entrance~\cite{mann,hughes,elkady}. These models were motivated by
observations of higher coke concentrations in the external surfaces of several
catalysts~\cite{rostrup,butt}, which is typical of systems with a parallel
mechanism of coke formation~\cite{froment}.
The common feature of most previous models is that the shape of the deposits is
defined a priori, independently of the particular features of the diffusion and
reaction processes. However, the interplay between transport of the
precursors, reactions and coke formation is a fundamental aspect for catalyst
deactivation.

The aim of this work is to study the morphology of the deposits
generated by a statistical model of diffusion and aggregation of a
chemical species inside a catalyst pore. The shape of a deposit changes in
time due to the diffusion-limited aggregation of residues. Two transport
regimes and various aggregation conditions will be considered, in order to
show the features of the deposits which
do not depend on the details of the diffusion and aggregation processes. The
model neglects the internal structure of the coke precursors (molecules or
radicals), considering a single diffusing/aggregated species, called particle
$C$, inside the pore. The transport is modeled by random walks,
considering separately the cases of molecular diffusion and Knudsen diffusion.
In the case of molecular diffusion, the possibility of a small diffusional
bias along the pores is analyzed. Details of the bond formation are also
simplified by considering simple stochastic rules for aggregation of the
diffusing species. Depending on the stochastic rule, dense or ramified deposits
are obtained.  Computationally, these models are similar to those of
diffusion-limited aggregation (DLA)~\cite{dla}, which have applications to
other fields, such as electro deposition~\cite{meakin,leger}. From this
point of view, the main differences are the restriction of transport processes
to a long pore of finite width, the possibility of aggregating the diffusing
species at the whole surface and the more complex probabilistic rules of
aggregation. 

Although we are not considering the details of a particular deactivation
process, we will derive conclusions on the large length scale features of the
deposits which are in qualitative agreement with some experimental
results~\cite{rostrup,butt}.
On the other
hand, they disagree with some of the most frequent theoretical assumptions for
the coke profiles in continuous models. Among the main conclusions, both for
Knudsen diffusion and some cases of molecular diffusion with small bias
along the pore (weak convection effects), we will show that the coke deposits
approximately have a wedge shape and plugging occurs at the pore mouth, after
a transient with uniform deposition. However, the coke concentration near the
pore mouth increases in time much faster than the concentration inside the
pore, in contrast to the assumption of a wedge shape of constant slope. It
suggests that incorporation of diffusion-limited aggregation mechanisms for
coking in models for real catalysts may be relevant for a reliable description
of their behavior. Another possibility is the incorporation of the effective
growth laws obtained here in continuous reaction-diffusion models which
require a prior assumption of the (time-dependent) shape for the coke deposits.

This paper is organized as follows. In Sec. II we will present the models of
diffusion and aggregation. In Sec. III we will discuss qualitative and
quantitative properties of the coke profiles generated by those models. In Sec.
IV we summarize the results and present our conclusions.

\section{Models}
\label{secmodel}

In order to simplify the porous structure of the catalyst and focus on the
shape of the coke deposits inside the pores, we will consider the diffusion and
aggregation processes at isolated pores. Each pore is a semi-infinite slab
with discretized positions, as shown in Fig. 1. Different values of the width
$W$ are considered and are expressed in lattice units. Such approximation
was previously adopted in several models of diffusion, reaction and
deactivation by coking~\cite{beyne,muhammad,hughes,elkady}.

\begin{figure}[ht]
\centering
\includegraphics[clip,width=0.80\textwidth, 
height=0.40\textheight,angle=0]{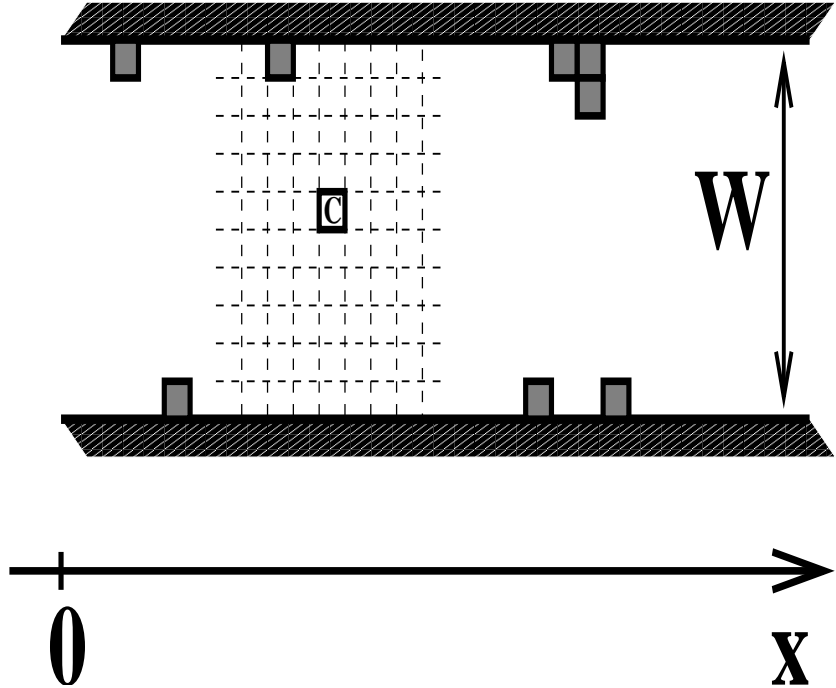}
\caption{Long pore of width $W$ with an internal square lattice structure and
entrance at $x=0$. The empty square is a diffusing particle $C$ which moves on
that network. Grey squares represent aggregated particles $C$.}
\label{fig1}                        
\end{figure}     

The coke precursors are represented by particles $C$, which enter the pore at
its left side (pore mouth $x=0$). The initial vertical position
of each incident particle is randomly chosen. These initial conditions apply
to all diffusion and aggregation models considered in this work. to all It is
assumed that particles $C$ were produced not only inside the pore which is
currently under study, but also in other regions of the reactor. Equivalently,
it is being assumed that the deposit at a certain pore is mainly formed by
residues produced elsewhere, which enter that pore through its opening.

The diffusion processes inside a pore will be simplified by restricting the
positions of particles $C$, diffusing or aggregated, to the sites of an
internal square lattice. Random walks on that network represent the molecular
diffusion regime and ballistic motion represents Knudsen diffusion, as
detailed below. Pore widths $W=10$, $W=20$ and $W=40$ (measured in lattice
units) were considered.

In Knudsen diffusion, the mean free path of molecules or radicals is much
larger than the width of
the pore. This low density regime inside a pore is typical of real catalytic
processes. In our model, each $C$ particle is
released at $x=0$ with velocity in a randomly chosen direction and
executes ballistic motions between collisions with the pore walls or
with the currently formed deposit (Fig. 2a). When the particle $C$ encounters a
pore wall or a previously aggregated particle, it may aggregate irreversibly
at that position according to the probabilistic rules I or II, described below.
If it does not aggregate, then it moves to a
lattice site in its neighborhood with velocity chosen in a random direction
and continues moving until colliding again. This random motion of a single
particle is executed until it aggregates. After aggregation,
a new particle $C$ is left at $x=0$. Since the pore is very long, the number
of particles $C$ that leave it at the right side during the simulations is
negligible.

The second regime considered here is that of molecular diffusion, in which
the mean free path of a particle $C$ is much smaller than the width of the
pore. In this case, particles $C$ execute biased random walks inside the pore
(Fig. 2b), in which steps in the $y$ direction (up or down in Fig. 2b)
and in the negative $x$ direction (left in Fig. 2b) are performed with rate
$1$, while steps in the positive $x$ direction (right in Fig. 2b) are
performed with rate $u$, with $u>1$. Consequently, $v\equiv u-1$ represents
the intensity of the diffusional bias. This diffusion mechanism is supposed to
describe the interaction of the particle $C$ with the remaining molecules of
the fluid which will not form the deposit and, consequently, do not need to be
explicitly represented in the model.

\begin{figure}[ht]
\centering
\includegraphics[clip,width=0.80\textwidth, 
height=0.40\textheight,angle=0]{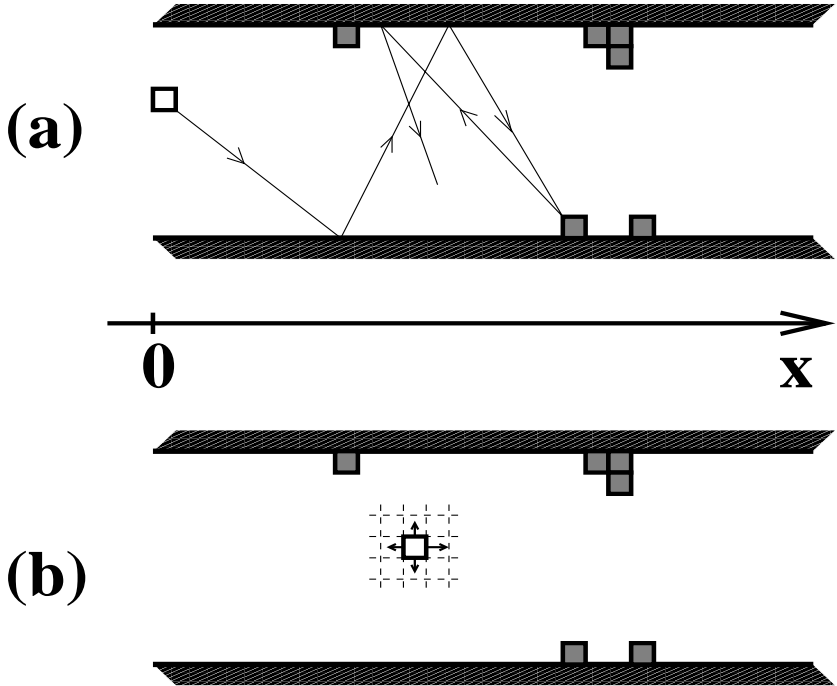}
\caption{(a) Illustration of the ballistic motion of a particle (empty square)
inside a catalyst pore in the model of Knudsen diffusion. Velocity
directions are random after collisions. Grey squares represent aggregated
particles. (b) Illustration of the biased random walk of a particle (empty
square) in the molecular diffusion model. The arrows indicate possible
movements and their lengths illustrate the bias along the positive $x$
direction.}
\label{fig2}                        
\end{figure}     

Now we describe the aggregation rules, both of them considered in simulations
of Knudsen and molecular diffusion. The sites at the pore walls and the
previously aggregated $C$ are treated equivalently as occupied sites. When a
particle $C$ is at a site with one or more occupied neighbors, it may
aggregate irreversibly at that position with probability $p_{ag}$.
Some possible aggregation positions are shown in Fig. 3, with the corresponding
numbers of occupied nearest neighbors, $n$. It is expected that $p_{ag}$
increases with $n$, but their relation strongly depends on
the specific catalyst and coke precursor. Thus, since our aim is to
investigate general features of coke deposits, we will consider two different
aggregation rules with different functional dependences of $p_{ag}$ on $n$.

\begin{figure}[ht]
\centering
\includegraphics[clip,width=0.80\textwidth, 
height=0.40\textheight,angle=0]{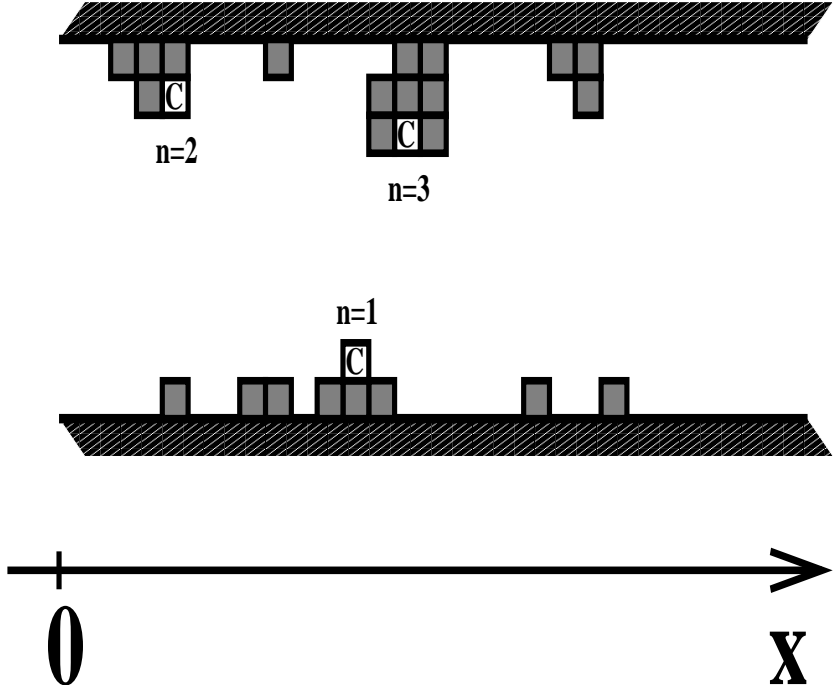}
\caption{Some possible aggregation positions for diffusing particles
(empty squares with label $C$) with the corresponding numbers of nearest neighbors
$n$.}
\label{fig3}                        
\end{figure}     

In the first aggregation rule, hereafter called rule I,
\begin{equation}
p_{ag} \equiv p^{\left( 4-n\right)} .
\label{rule1}
\end{equation}
In our simulations, we will consider the cases $p=0.1$ and $p=0.05$. The
value of the parameter $p$ is the aggregation probability for a particle $C$
with three aggregated neighbors. With the above values of $p$,
$p_{ag}$ rapidly increases with $n$:
$p_{ag}\sim {10}^{-2}$ with two neighbors and $p_{ag}\sim{10}^{-3}$
with one neighbor. Thus, while aggregation to a flat surface is very
difficult ($p_{ag}\sim{10}^{-3}$), it is much easier at
irregular regions of the deposit, in which the diffusing particle may
encounter two or three neighbors.

In the second aggregation rule, hereafter called rule II,
\begin{equation}
p_{ag} \equiv 1-q^n .
\label{rule2}
\end{equation}
The values $q=0.99$ and $q=0.995$ will be considered in our simulations. This
aggregation rule represents a much slower decrease of $p_{ag}$ when $n$
decreases; typically, $p_{ag}$ doubles from $n=1$ to $n=2$ and increases
$50\%$ from $n=2$ to $n=3$. Rule II may represent the
effects of adsorption and subsequent desorption of a particle $C$ when it is
near the solid deposit. In this case, the parameter $q$ is of order
$\exp{\left( -E/k_BT\right) }$, where $E$ is an activation energy and $T$ is the
temperature.

For any set of the model parameters, the average values were taken over $100$
to $1000$ realizations. 

DLA with aggregation rules similar to rule I (Eq. \ref{rule1}) were
previously analyzed by various authors~\cite{banavar,aukrust}, leading to a
crossover from compact to ramified clusters. DLA models in striped geometries
were previously studied in Ref. \protect\cite{meakin1}, but the aggregation
probabilities were independent of the number of neighbors (the usual condition)
and the particle flux was very different from that considered here.

\section{Results}
\label{secresults}

First we discuss the results obtained for Knudsen diffusion,
i. e. with ballistic motion of species $C$ inside the pore. There is no bias
in this case.

In Fig. 4a we show a sequence of configurations of the deposit
inside a pore of width $W=40$, obtained in simulations of rule I with $p=0.1$.
In Fig. 4b we show a sequence of configurations obtained in simulations of
rule II with $q=0.99$. Due to the probabilistic nature of the process, there
are height fluctuations around any position $x$ along the pore.
There is an initial period of uniform deposition. Subsequently, if the height
is averaged over lengthscales of order $W$ or larger, in both cases this
average value decreases with $x$. At those lengthscales, the deposit has an
approximately wedge shape, deviating from this form only near the entrance
when the pore is almost plugged. At this critical regime, the coke
concentration near the entrance increases much faster than at the inner part
of the pore. Also notice that the deposits obtained with rule I are compact,
while the deposits obtained with rule II have a ramified structure.

\begin{figure}[ht]
\centering
\includegraphics[clip,width=0.80\textwidth, 
height=0.40\textheight,angle=0]{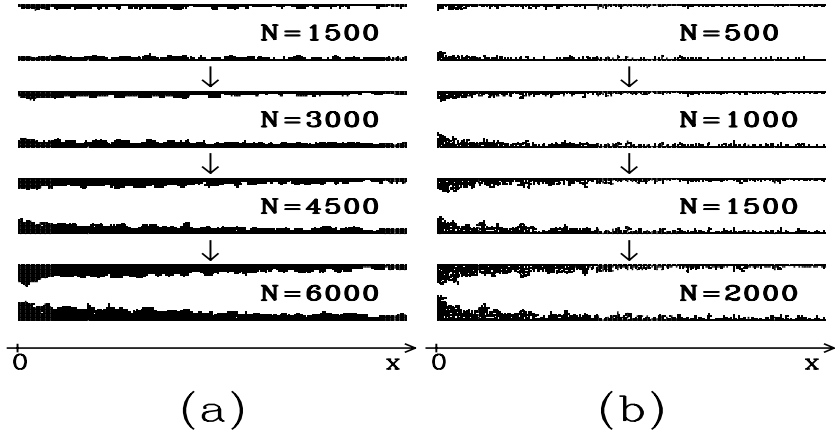}
\caption{Sequence of configurations of the left side of a pore with $W=40$
for Knudsen diffusion: (a) aggregation rule I with
$p=0.1$; (b) aggregation rule II with $q=0.99$. The number of deposited
particles of each configuration, $N$, is indicated.}
\label{fig4}                        
\end{figure}     

For other values of parameters $p$ and $q$, similar results were obtained. In
any case, the pores plug at their entrances, as expected from the mechanism
of particle injection at that region. Indeed, higher coke
concentrations at the external surface of catalyst particles (pellets) is 
experimentally observed in several situations~\cite{rostrup,butt}.
However, before discussing other nontrivial features of the Knudsen diffusion
model, we will show that similar results are obtained in the
molecular diffusion regime, even with a small bias along the pore.

For the molecular diffusion model, several values of $v$ were considered
(typically between $v=0$ and $v=10.0$) for each value of parameters $p$
(aggregation rule I) and $q$ (rule II). Small $v$ represents weak diffusional
bias, while large $v$ represents a rapid flux along the pore, with rare
collisions with its surface. Large $v$ is certainly not a realistic
description of real catalytic processes, but it was also analyzed for
completeness.

In Fig. 5a we show a sequence of configurations of a pore of width $W=40$,
obtained in simulations of rule I with $p=0.1$ and $v=0.1$. In Fig. 5b we show
a sequence of configurations of the same pore using rule II with
$q=0.99$ and $v=0.2$. We also observe the decrease of the average height of the
deposit along the pore at lengthscales of order $W$ or larger, which is
confirmed for the other values of $p$ and $q$ and small $v$.

\begin{figure}[h]
\centering
\includegraphics[clip,width=0.80\textwidth,
height=0.40\textheight,angle=0]{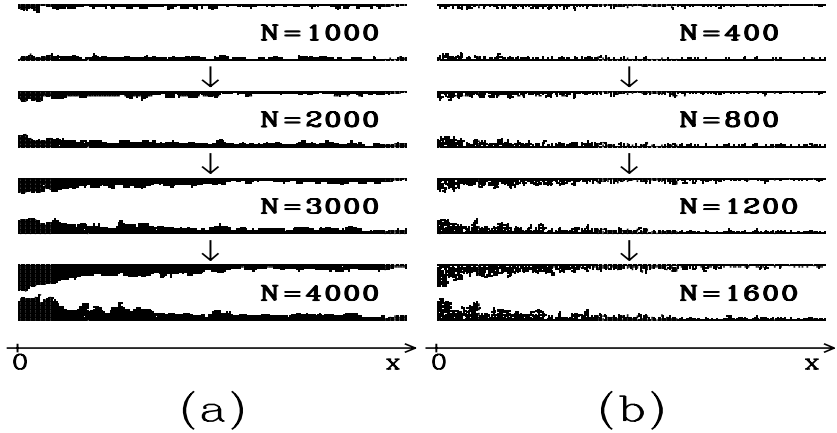}    
\caption{Sequence of configurations of the left side of a pore with $W=40$
for particles executing biased molecular diffusion: (a) aggregation rule I with
$p=0.1$ and $v=0.1$; (b) aggregation rule II with $q=0.99$ and $v=0.2$. The
number of deposited particles of each configuration, $N$, is indicated.
}
\label{fig5}
\end{figure}  

When the diffusional bias $v$ along the pore increases, the typical morphology
of the deposit changes, showing much larger heights' fluctuations. The pore
plugs at positions far from $x=0$ and the approximately wedge shape
of the deposit disappears.

In order to analyze quantitatively the conditions of pore plugging in the
molecular diffusion model, we calculated
the average horizontal position $\langle x_p\rangle$ at which the pores
become plugged, i. e. the average position of the last aggregated particle.
In Fig. 6a we show $\langle x_p\rangle$ versus the bias $v$ for aggregation
rule I, with two values of $p$ and three different widths. In Fig. 6b we show
$\langle x_p\rangle$ versus $v$ for aggregation rule II, with two values of
$q$, for $W=20$ and $W=40$. In all cases, there is a range of small $v$ in
which $\langle x_p\rangle$ is significantly smaller than the widths $W$: $v\leq
0.02$ for rule I with $p=0.1$ and $p=0.05$, $v\leq 0.1$ for rule II with
$q=0.99$ and $q=0.995$. It indicates plugging at the pore mouth, which is
confirmed by visual inspection of various pore configurations (see, for
instance, Figs. 5a and 5b). The qualitative features of the deposits with no
bias $v=0$ are the same (saturation of $\langle x_p\rangle$ at a small value).

\begin{figure}[h]
\centering
\includegraphics[clip,width=0.80\textwidth,
height=0.40\textheight,angle=0]{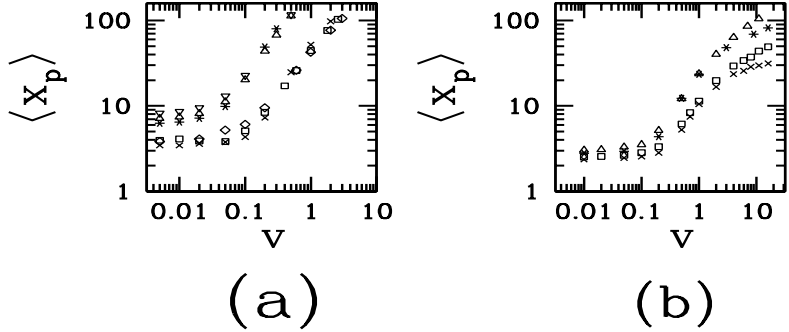}    
\caption{Average horizontal position at which the pores plug, $\langle
x_p\rangle$, as function of the diffusional bias $v$, for the molecular
diffusion model. (a) Aggregation rule I with: $p=0.1$
and $W=10$ (diamonds), $W=20$ (squares) and $W=40$ (crosses); $p=0.005$ and
$W=10$ (down triangles), $W=20$ (up triangles) and $W=40$ (asterisks). (b)
Aggregation rule II with: $q=0.99$ and $W=20$ (squares) and $W=40$ (crosses);
$q=0.995$ and $W=20$ (triangles) and $W=40$ (asterisks).
}
\label{fig6}
\end{figure}  

Now we consider some quantitative properties of the deposits generated by both
diffusion models.

One of the parameters that characterize the shape of the deposits is the
average horizontal position of aggregation, $\langle x_{ag}\rangle$,
which is the position of the center of mass relative to the entrance. This
quantity is useful for describing their coarse grained features,
independently of local thickness fluctuations. $\langle x_{ag}\rangle$ was
measured as a function of the number of deposited particles $N$, which is
proportional to the mass of the deposit.

In Fig. 7 we show ${\langle x_{ag}\rangle}/W$ as a function
of $N$ for the model of Knudsen diffusion
with both aggregation rules. Average values were taken only over the set of
configurations which were not blocked. Moreover, Fig. 7 shows only data for
values of $N$ in which more than $50\%$ of the pores were not blocked.
Consequently, they represent the growth regime of the deposits but not
the critical pore blocking.

\begin{figure}[h]
\centering
\includegraphics[clip,width=0.80\textwidth,
height=0.40\textheight,angle=0]{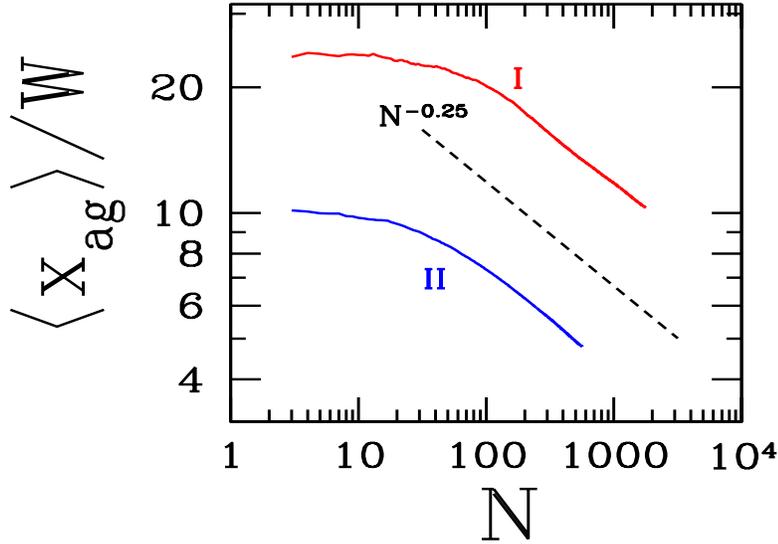}    
\caption{Ratio between the average horizontal position of
aggregation $\langle x_{ag}\rangle$ and the pore width $W$ as a function of
the number of deposited particles $N$, for the model of Knudsen diffusion:
aggregation rule I with $W=20$ and $p=0.1$;
aggregation rule II with $W=20$ and $q=0.99$. The dashed
line illustrates the decay as $N^{-0.25}$.
}
\label{fig7}
\end{figure}  

Two regimes are observed in Fig. 7. The deposition initially occurs in a large
region of the pore and not only at its entrance - the entrance is typically the
region with ${{\langle x_{ag}\rangle}/W} <1$. Subsequently, the deposit becomes
large, with significant thickness fluctuations that reduce the effective
diffusivity and makes it easier for particles $C$ to aggregate to the
irregularities of the surface. Consequently, $\langle x_{ag}\rangle$ slowly
decreases with $N$. The results for both models indicate a decay $N^{-\beta}$,
with $\beta\sim 0.25$, as illustrated in Fig. 7.
This relation is just an effective form of decay for restricted ranges of $N$,
since the local slopes of the curves in Fig. 7 have significant fluctuations
in the regime of $\langle x_{ag}\rangle$ decay (mainly for model II). For
finite and small widths, as expected in catalyst pores, this limitation in
the range of $N$ is natural. Consequently, that form of decay may not be
interpreted as an universal scaling law for such processes within the
expected limits of application.

In Fig. 8, we show the results for the molecular diffusion model in the small
$v$ regime.

In the case of aggregation rule I, the evolution of $\langle
x_{ag}\rangle$ is similar to that of the Knudsen model, also including a
decay with $\beta\sim 0.25$ for large deposits. For much smaller bias
(including $v=0$), the values of $\langle x_{ag}\rangle$ are much smaller
and its decrease is slower, which corresponds to effective exponents $\beta$
slightly smaller ($\beta\sim 0.15$). For
aggregation rule II, with a relatively large bias, $\langle x_{ag}\rangle$
also decays very slowly until the pore blocking regime. This is an effect of
the weak dependence of $p_{ag}$ on the number of aggregated neighbors $n$ for
this rule (see Sec. II) and of the biased diffusion, both reducing the
influence of the irregularities on the aggregation process. 

\begin{figure}[h]
\centering
\includegraphics[clip,width=0.80\textwidth,
height=0.40\textheight,angle=0]{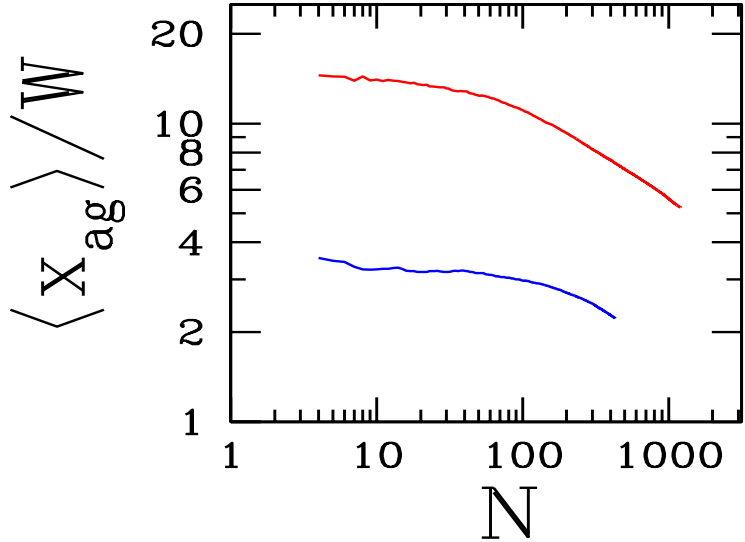}    
\caption{Ratio between the average horizontal position of
aggregation $\langle x_{ag}\rangle$ and the pore width $W$ as a function of
the number of deposited particles $N$, for the molecular diffusion model with
small bias: aggregation rule I with $W=20$, $p=0.1$ and $v=0.05$ (top curve);
aggregation rule II with $W=20$, $q=0.99$ and $v=0.2$ (lower curve).
}
\label{fig8}
\end{figure}

The above results show that, in the model of Knudsen diffusion, which is
expected to be more adequate for real catalytic processes, the deposits
generated by diffusion and aggregation of particles entering the pore have
some features which do not depend on the details of the aggregation
conditions. For small deposits, the assumption of a uniform coke profile
($\langle x_{ag}\rangle\approx const$) is reasonable; indeed this
approximation is frequently adopted in continuum models. However, as the
deposit becomes larger, neither the uniform profile nor the assumption of a
wedge-shaped deposit of constant slope~\cite{mann,hughes,elkady} is supported
by our microscopic models. The wedge-shaped deposit of constant slope would
correspond to a linear increase of $\langle x_{ag}\rangle$ with $N$, in
contrast to the decrease shown in Figs. 7 and 8. Instead, the assumption of
the above effective $N^{-\beta}$ law with small $\beta$ ($\beta\sim 0.25$ for
Knudsen diffusion) seems to be a better approximation and might eventually be
tested in continuous models.

\section{Conclusion}
\label{secconclusion}

We presented results of simulations of various models of transport and
aggregation of a single species in isolated pores in order to represent
the basic mechanisms of coke deposition in catalyst pores. In the first model,
representing the Knudsen regime, the diffusing species (particles $C$)
executed ballistic motion between collisions with the pore walls or with the
deposit. In the second model, suitable for the molecular diffusion regime, the
$C$ particles executed biased random walks inside the pores. Two probabilistic
rules for aggregation to the walls or to the deposit were considered, leading
to dense or ramified deposits.

In the Knudsen diffusion model or in the molecular diffusion case with
weak bias along the pore, the
deposits formed at the pore surface have approximately wedge shapes at
lengthscales of order $W$ or larger. Consequently, the pore plugging occurs at
the pore mouth. The average position of aggregation slowly decreases with
the number of particles, approaching the pore mouth ($x=0$) as the deposit
becomes larger. These results for Knudsen diffusion do not depend on the
details of the aggregation mechanisms. An effective decay ${\langle
x_{ag}\rangle}\sim N^{-0.25}$ is obtained after a transient regime of uniform
deposition and before the critical regime of pore plugging. Similar results
are found in some conditions of the molecular diffusion models with small
bias. Consequently, the hypothesis of uniform coke profiles along the
pores~\cite{beyne,muhammad} and the hypothesis of a wedge-shaped deposit of
constant slope~\cite{mann,hughes,elkady}, frequently adopted in continuous
reaction-diffusion models during the whole deposition period, are not
supported by our microscopic models.

Those models were not designed to describe
quantitatively any real catalyst because our interest was to analyze
some general features of the deposits inside the pores. On the other hand,
since very different conditions of diffusion and
aggregation were considered, our results suggest that the above features are
not consequences of a particular model. Instead, they are general conclusions
that follow from the assumption that coke deposition is a consequence of
diffusion-limited aggregation of carbonaceous residues inside catalyst pores.
Thus, we expect that the information obtained from our study can be applied to
other processes, possibly of technological interest. For instance, in models
with prior assumptions of the shapes of coke profiles, the approximate
dependence of ${\langle x_{ag}\rangle}$ on the mass of the deposit might be
tested. The extension for modeling coke deposition in a specific catalyst
system would also include the description of diffusion and chemical reactions
of several species inside the pores, as well as considering their connectivity.

\vskip 1cm

{\bf Acknowledgements}

This work was partially supported by CNPq and FAPERJ (Brazilian agencies).


\end{document}